\documentclass[aps,superscriptaddress,showpacs,nofootinbib]{revtex4}
\usepackage{graphics}
\usepackage{bm}

\input{epsf}

\newcommand{\be}{\begin{equation}}
\newcommand{\ee}{\end{equation}}
\newcommand{\bea}{\begin{eqnarray}}
\newcommand{\eea}{\end{eqnarray}}

\begin{document}

\preprint{UG-FT-191/05}

\preprint{CAFPE-61/05}

\title{Sneutrino warm inflation in the minimal supersymmetric model}
\author{Mar Bastero-Gil}
\email{mbg@ugr.es}
\affiliation{Departamento de F\'{\i}sica Te\'orica y del Cosmos,
  Universidad de Granada, Granada-18071, Spain}

\author{Arjun Berera}
\email{ab@ph.ed.ac.uk}
\affiliation{ School of Physics, University of
Edinburgh, Edinburgh, EH9 3JZ, United Kingdom}

\begin{abstract}
 
The model of RH neutrino  fields coupled to the MSSM is shown
to yield a large parameter regime of warm inflation.
In the strong dissipative regime, it is shown that
inflation, driven by a single sneutrino field, occurs
with {\it all field amplitudes below the Planck scale}.
Analysis is also made of leptogenesis, neutrino mass generation
and gravitino constraints.
A new warm inflation scenario is purposed in which one
scalar field drives a period of warm inflation and
a second field drives a subsequent phase of reheating.
Such a model is able to reduce the final temperature after 
inflation, thus helping to mitigate gravitino constraints.

\medskip                                   

\noindent
keywords: cosmology, inflation
\end{abstract}
                                                                                

\maketitle

\section{Introduction}
\label{sect1}

In recent times, the idea of inflation being driven by
the bosonic supersymmetric partner to a neutrino field has
generated interest \cite{csnu,hsnu}.  The idea is not new
\cite{snuoriginal,snusugra}, 
but impetus has been gained after the experimental discovery
of neutrino masses and mixing and an explanation through
the seesaw mechanism \cite{seesaw}.
In supersymmetric realizations of the seesaw mechanism, the right-handed
neutrinos have bosonic partners, sneutrinos, which are
singlet fields, thus possible inflaton candidates.
Model building typically proceeds by simply adding on
the additional right-handed neutrino fields to an
existing model. Thus the simplest supersymmetric
model that emerges is an extended version
of the MSSM, with now three families of right-handed
neutrinos added on.

Two types of sneutrino inflation models have been examined, 
chaotic \cite{csnu} and hybrid \cite{hsnu} sneutrino inflation.
The chaotic model is the simplest to construct, since all
it requires is a monomial potential which can easily be
obtained directly from the sneutrino fields.
However this model suffers from the large field problem,
in that the sneutrino field that drives inflation will have
to have a field amplitude above the Planck scale.
In the effective field theory interpretation of global Supersymmetric
models, they are regarded as low-energy limits of some more complete
supergravity (sugra) theory. However, for example in ``minimal'' sugra 
the exponential factor in front of the potential would prevent any
scalar field from getting a value larger than $m_P$. Chaotic inflation
would be possible with other more involved choices of the Khaler
potential \cite{snusugra} such that sugra corrections are kept under
control. Still, in general in these models there are an infinite number of  
non-renormalizable operators suppressed by the 
Planck scale.
As such, once the field amplitude exceeds this scale, 
an infinite number of parameters would require fine-tuning,
so leaving no predictability in the theory.
It is for this reason that chaotic inflation models are
not amenable to particle physics model building.
Hybrid inflation scenarios overcome the large field problem,
since all field amplitudes are well below the Planck scale.
However for sneutrino inflation, these models require introducing 
two additional superfields aside from the
the right-handed neutrino fields \cite{hsnu}.  As such, this model is
more contrived than the chaotic model.
Nevertheless, up to now the hybrid model appears to be
the simplest model in which to implement sneutrino inflation
and be amenable to particle physics model building.

In this paper an even simpler model of sneutrino inflation
is presented.   In particular we show that monomial
potentials, which can be constructed with only
the right-handed sneutrino fields, when coupled to
the MSSM, realize warm inflationary regimes.
We show that in such regimes, due to the effect of
strong dissipation, the field amplitudes of all sneutrino
fields are well below the Planck scale, thus allowing
such models to be consistent with particle physics model building.
   
The paper is organized as follows.  The basic model is presented
in Sect. \ref{sect2}.  The dissipative effects and basic equations
of warm inflation for this model are obtained in Sect. \ref{sect3}.
The results of the sneutrino warm inflation scenario, which incorporates
leptogenesis, are
given in Sect. \ref{sect4}.  A issue that emerges in Sect. \ref{sect4}
is that the final temperature after inflation is too large
to adequately control gravitino constraints.  To
improve this situation, in Sect. \ref{sect5} a
new warm inflation scenario is presented.
In Sect \ref{sect6} neutrino mass generation from
this scenario are examined.
Finally in Sect. \ref{sect7} we summarize our results.

\section{Model}
\label{sect2}

We consider the model of
three generations of right-handed neutrinos, $N_i$, coupled to the MSSM
with the superpotential
\be
W= \frac{M_i}{2} N_i N_i+ (h_N)_{ij} H_u L_i N_j+ (h_L)_{ij} H_d L_i
E^c_j + (h_u)_{ij} H_u Q_i U^c_j + (h_d)_{ij} H_d Q_i D^c_j \,.
\label{snumssm}
\ee
The above model contains all the usual MSSM matter superfields,
$H_u$, $H_d$ the Higgs doublets giving masses to the
up and down quarks respectively, $Q_i$, the left-handed quarks,
$U_i$, $D_i$, the right-handed up and down quarks respectively, and
$L_i$, $E_i$ the left-handed  and right-handed leptons.                                                               
During inflation, assuming that at least one of the sneutrinos has got
a non zero vacuum expectation value (vev), the relevant terms in the
potential are: 
\be
V= M_i^2 |N_i|^2+ |(h_N)_{ij}|^2|L_i H_u|^2+ 2 Re [(h_N^*)_{ij} M_i
  N_j L_i^* H_u^*]+ |(h_N)_{ij}|^2(|H_u|^2+ |L_i|^2)|N_j|^2  + \cdots \,.
\ee
For a large value of $N_i$ we do not have to worry about soft susy
breaking terms, and then all the spectrum remains massless (
$H_u=H_d=0$), except for the fields that couple directly to the
sneutrinos, i.e., $H_u$ and the lepton doublets $L_j$. With $\langle
N_i\rangle = \phi_{Ni}/\sqrt{2}$, the scalars for example get masses
\be
m_{H_u}^2 = \frac{1}{2} \sum_{j}h_{Nj}^2 \phi_{Nj}^2 \,
\ee
and similarly for the sleptons, where
\be
h_{N_j}^2 = \sum_{i}|(h_N)_{ij}|^2 \, .
\label{yukawahN}
\ee
                                                                                
\section{Dissipative inflationary dynamics}
\label{sect3}

The interaction of the inflaton with other fields leads in general
not only to modifications of the inflaton effective potential,
but also to dissipative effects \cite{br05,br}.  
These effects result in radiation 
production during inflation as well as modify the inflaton evolution
equation with energy non-conserving terms.  If these dissipative
effects are adequately large, they can alter the standard picture of
inflation, leading to warm inflation \cite{wi}.

An analysis of various interaction configuration \cite{br05,br} 
has shown that warm
inflation occurs generically in many typical
inflaton models.
For example recently we showed in \cite{bb1} that the popular
SUSY hybrid inflation model has a sizable parameter regime of
warm inflation.  In this section we show that warm
inflation occurs in the sneutrino-MSSM model Eq. (\ref{snumssm}).

A basic interaction structure that has been shown 
in \cite{br05,br,Hall:2004zr}
to produce sizable dissipative effects has the form of
the bosonic inflaton field coupled to a heavy bosonic field
which in turn is coupled to a light fermionic field.
Such a structure can easily be identified in the sneutrino-MSSM
model Eq. (\ref{snumssm}).  For this consider the simplest case
where only one sneutrino dominates the energy density during
inflation, say $N_1$, thus acting the role of the inflaton field.
Then from Eq. (\ref{snumssm}) the following relevant interaction
configuration can be extracted
\begin{equation}
{\cal L}_I = - |h_N|^2 |N_1|^2|H_u|^2 + 
h_{t} H_u \bar t_R t_L + h.c. ,
\end{equation}
thus the inflation $N_1$ couples to the up Higgs field, and
for a large amplitude for $N_1$, the $H_u$ field then becomes
heavy.  This Higgs field in turn is coupled to the top fermion
fields which are massless during inflation.
Dissipative effects occur because as the inflaton amplitude changes, it
implies a change to the $H_u$ mass.  This results
in a coherent excitation of the $H_u$ field, which then
decays into the light top fermions with decay rate
\begin{equation} 
\Gamma_{t}= \frac{3h_t^2}{16 \pi} m_{H_u} \,.
\label{decay}
\end{equation} 
From the dissipative calculations
in Refs. \cite{br05,br} this sort of interaction leads to the effective
inflaton evolution equation
\be
\ddot \phi + (3 H + \Upsilon_N) \dot \phi + V^\prime =0 \,.
\label{ddotphi}
\ee
where the dissipative coefficient, based on the results in
\cite{br05,br}, can be determined to be
\begin{equation}
\Upsilon_N \simeq \frac{\sqrt{\pi}}{20} Y_N^{3/2} Y_t \phi_N \,,
\label{disscoef}
\end{equation}
with $Y_N \equiv h_N^2/(4 \pi)$ and  $Y_t ={h_t^2}/{4 \pi}$.
Also in Eq. (\ref{ddotphi})
the potential and the Hubble rate are
\begin{equation}
V \simeq \frac{1}{2}M_N^2 \phi_N^2 \
\end{equation}
and
\begin{equation}
H^2 \simeq \frac{M_N^2 \phi_N^2}{6 m_P^2} \,.
\end{equation}

The dissipative term in Eq. (\ref{ddotphi}) leads to radiation production
which in the expanding spacetime obeys the equation
\be
\dot \rho_R + 4 H \rho_R = \Upsilon_N \dot \phi_S^2 \,.
\label{dotrhor}
\ee
Although the basic idea of interactions leading to dissipative effects
during inflation is generally valid, the above set of equations
has strictly been derived in \cite{br05,br}
only in the adiabatic-Markovian
limit, i.e., when the fields involved are moving slowly, which
requires 
\be
\frac{\dot \phi}{\phi} < H < \Gamma_t \,,
\label{markovian}
\ee
with $\Gamma_t$ being the decay rate Eq. (\ref{decay}). 
The second inequality, $ H < \Gamma$
is also the condition for the radiation (decay products) to thermalize.   

Thus in general any inflation model could have two very distinct 
types of inflationary dynamics, which have been termed cold and
warm \cite{wi,br05,br}.  The cold inflationary regime
is synonymous with the standard inflation picture
\cite{oldi,ni,ci}, in which dissipative effects are completely
ignored during the inflation period.  On the other hand,
in the warm inflationary
regime dissipative effects play a significant role in 
the dynamics of the system.  A rough quantitative measure
that divides these two regimes is
$\rho_R^{1/4} \approx H$, where
$\rho_R^{1/4} > H$ is the warm inflation regime 
and $\rho_R^{1/4} \stackrel{<}{\sim} H$
is the cold inflation regime. This criteria is independent of
thermalization, but if such were to occur, one sees this criteria
basically amounts to the warm inflation regime corresponding to when $T
> H$. This is easy to understand since the typical inflaton mass during
inflation is $m_\phi \approx H$ and so when $T>H$, thermal fluctuations
of the inflaton field
will become important. This criteria for entering the warm inflation
regime turns out to require the dissipation of a very tiny fraction of
the inflaton vacuum energy during inflation. {}For example, for
inflation with vacuum ({\it i.e.} potential) energy at the GUT scale
$\sim 10^{15-16} {\rm GeV}$, in order to produce radiation at the scale
of the Hubble parameter, which is $\approx 10^{10-11} {\rm GeV}$, it just
requires dissipating one part in $10^{20}$ of this vacuum energy density
into radiation. Thus energetically not a very significant amount of
radiation production is required to move into the warm inflation regime.
In fact the levels are so small, and their eventual effects
on density perturbations and inflaton evolution are so significant, that
care must be taken to account for these effects in the analysis of any
inflation models.

The conditions for slow-roll inflation ($ \dot \phi_S^2 \ll V$, $\ddot
\phi_S \ll H \dot \phi_S$) are modified in the
presence of the extra friction term $\Upsilon_N$, and we have now:
\bea
\epsilon_\Upsilon &=& \frac{\epsilon_H}{(1+r)^2}
\simeq \frac{2}{C_\Upsilon^2}\left(\frac{M_N}{\phi_N}\right)^2 
\label{epsups}\,,\\
\eta_\Upsilon &=& \frac{\eta_H}{(1+r)^2}
\simeq \frac{2}{C_\Upsilon^2}\left(\frac{M_N}{\phi_N}\right)^2 
\label{etaups} \,,
\eea
where
\begin{equation} 
r \equiv \frac{\Upsilon_N}{3 H} =  
\frac{\sqrt{6 \pi}}{60} Y_N^{3/2} Y_t
\frac{m_P}{M_N}= C_\Upsilon \frac{m_P}{M_N} \,.
\label{eqr}
\end{equation} 
and $\epsilon_H \equiv m_P^2 V'^2/(2V^2)$, 
$\eta_H \equiv m_P^2 V''/V$ are 
the standard cold inflation slow-roll parameters,
in which there are no dissipation effects.
In the slow-roll regime, when
$\eta_\Upsilon < 1$ and $\epsilon_\Upsilon < 1$, 
Eqs. (\ref{ddotphi}) and (\ref{dotrhor})
are well approximated by:
\bea
\dot \phi_S &\simeq& -\frac{V^\prime}{3 H}\frac{1}{1+r}\,, \\
\rho_R &\simeq& \frac{\Upsilon_N}{4H} \dot \phi_S^2 \simeq
\frac{1}{2}\frac{r}{(1+r)^2} \epsilon_H V\,,
\eea
and the number of e-folds is given by:
\be
N_e \simeq - \int_{\phi_{Si}}^{\phi_{Se}} \frac{3 H^2}{ V^\prime} (
1 + r) d \phi 
\simeq \frac{(1+r)}{4m_P^2}
\left(\phi_{Ne}^2 - \phi_{end}^2 \right)  \,,
\label{Neups} 
\ee
where $\phi_{Ne} (\phi_{end})$ is the value of the field at 60 e-folds
(end of inflation). Inflation ends when
$\epsilon_\Upsilon \simeq
\eta_\Upsilon\simeq 1$ or when $\rho_R \simeq \rho_N$, whatever
happens first. In the former case we have 
$\phi_{end}^2\simeq 2 m_P^2/(1+r)^2$, whereas for $\rho_R \simeq
\rho_N$ then $\phi_{end}^2\simeq m_P^2 r/(1+r)^2$. In either case,
taking $r \gg 1$, we get 
\be
\phi_{Ne} \simeq \sqrt{\frac{4 N_e}{1+r}} m_P\,.
\ee
If we also want to keep the field below the Planck scale, we need $r >
4 N_e \simeq 240$. From Eq. (\ref{eqr}), taking $|h_N| \simeq 1$, this
gives the upper bound on the sneutrino mass $M_N \agt 8\times 10^{12}$
GeV.

The effect of the dissipative term, in addition to
producing a friction term for the inflaton field,  
leads to radiation production which can alter
density perturbations.
Approximately, one can say that when the radiation production leads to
$T > H$, the fluctuations of the inflaton field are induced by the
thermal fluctuations, instead of being vacuum fluctuations, with a
spectrum proportional to the temperature of the thermal bath. 
We notice that having $T>H$ does not necessarily require $\Upsilon_N >
3H$. Dissipation may  not be strong enough to alter the dynamics of the
background inflaton field, but it can be enough even in the weak
regime to affect its fluctuations, and therefore the
spectrum. Depending on the different regimes,
the spectrum of the
inflaton fluctuations $P^{1/2}_{\delta \phi}$ is given for
cold inflation \cite{Guth:ec}, weak
dissipative warm inflation \cite{Moss:wn,Berera:1995wh},  
and strong dissipative warm inflation \cite{Berera:1999ws}
respectively by
\bea
T < H:  && P^{1/2}_{\delta \phi}|_{T=0}\simeq \frac{H}{2 \pi} \,, \\
\Upsilon_N < H < T: && P^{1/2}_{\delta \phi}|_{T}
 \simeq \sqrt{TH} \sim
\sqrt{\frac{T}{H}}P^{1/2}_{\delta \phi}|_{T=0} \,, \\ 
\Upsilon_N > H: && P^{1/2}_{\delta \phi}|_{\Upsilon} \simeq
\left(\frac{\pi \Upsilon_N}{ 4 H}\right)^{1/4}\sqrt{T H}\sim 
\left(\frac{\pi \Upsilon_N}{ 4 H}\right)^{1/4}\sqrt{\frac{T}{H}}
P^{1/2}_{\delta \phi}|_{T=0} \,,   
\label{pphir}
\eea
with the amplitude of the primordial spectrum of the curvature
perturbation given by:
\be
P^{1/2}_{\cal R} = \left|\frac{H}{\dot
  \phi_S}\right| P^{1/2}_{\delta \phi} \simeq
\left|\frac{3 H^2}{V^\prime}\right| (1+r) P^{1/2}_{\delta \phi}
\label{spectrumr} \,.
\ee
Given the different ``thermal'' origin of spectrum, the spectral index
also changes with respect to the cold inflationary scenario
\cite{arjunspectrum,hmb1,warmspectrum,warmrunning}, even in the weak
dissipative warm inflation regime when the evolution of the inflaton field is
practically unchanged. 
General expressions for the spectral index are given in \cite{bb1}
and those relevant to the model in this paper will be given in the
sections that follow.

\section{Sneutrino warm inflation}
\label{sect4}

\begin{figure}[t] 
\hfil\scalebox{0.5} {\includegraphics{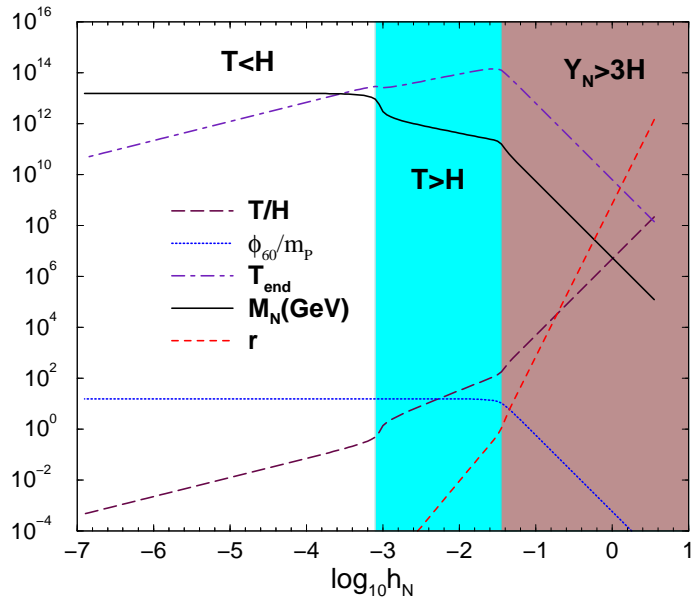}}\hfil
\caption{Values of the parameters $M_N$, $r$ and $\phi_{60}$ with respect
to the sneutrino Yukawa coupling $|h_N|$, 
for cold inflation ($T < H$), weak
dissipative regime ($T > H$, $\Upsilon_N < 3H$), 
and strong dissipative regime ($T, \Upsilon_N/3 > H$).}
\label{plot1}
\end{figure}

As we have seen, depending on the value of the dissipative coefficient
$\Upsilon_N$, and therefore that of the ratio $r=\Upsilon_N/(3H)$, we can
have standard cold inflation or warm inflation (with weak or strong
dissipation). In sneutrino inflation with the minimal matter content
of the MSSM plus 3 generations of RH (s)neutrinos
as shown in Eq. (\ref{snumssm}), there is a
well define dissipation channel during inflation due to the coupling
of the RH sneutrinos to the Higgs $H_u$, and the coupling of $H_u$ to
the top sector. From the experimental value of the top quark mass,
$m_t = 174 (178)$ GeV \cite{mtop}, 
the value of the top Yukawa coupling $h_t$ at the electroweak scale has to be
close to one, with $m_t \simeq h_t(m_t) v \sin \beta$, $v= \sqrt{v_u^2
+ v_d^2 }= 174$ GeV being the Higgs
vacuum expectation value (vev), and $\tan \beta=\langle H_u
\rangle/\langle H_d \rangle$ the ratio of the Higgs vevs. The top
Yukawa coupling increases due to the running 
with the scale, and depending on the  value of $\tan \beta$ it can
reach the perturbative bound 
$h_t(M_X) \simeq \sqrt{4 \pi}$ at
the unification scale $M_X$. Thus, although slightly model dependent,
its value at the inflationary scale will be in the range $[1, \sqrt{4
  \pi}]$. Then, without lost of generality we
can take its value  close to the perturbative bound $Y_t=h_t^2/(4
\pi) \simeq 1$, so that the dissipative coefficient 
Eq. (\ref{disscoef})  becomes
$\Upsilon_N \approx \frac{\sqrt{\pi}}{20} Y_N^{3/2} \phi_N$.
The free parameters in the model are then the
sneutrino-inflaton mass $M_N$ and its Yukawa coupling $|h_N^2|$ defined in
Eq. (\ref{yukawahN}). Imposing the COBE normalization on the
amplitude of the 
primordial spectrum of perturbations \cite{COBE,WMAP} generated during
inflation, we can fix one of these parameters, say the mass $M_N$, 
as dependent on the value of the coupling $h_N$.
This is plotted in Fig. (\ref{plot1}), where we
have included in addition to the value of $M_N$(GeV) the values of the
dissipative ratio $r=\Upsilon_N/(3H)$, 
the field value in $m_P$ units at 60 e-folds,
the temperature of the thermal bath at the end of inflation $T_{end}$,
and the ratio  $T/H$ during inflation. We can clearly distinguish the different
regimes in the plot depending on the sneutrino Yukawa value. For very
small values $|h_N| < 10^{-3}$ we recover the standard cold inflation
predictions, with $\phi_N > m_P$ and $M_N \simeq 2 \times 10^{13}$
GeV. In this regime, the Yukawa coupling plays no role during 
inflation, and the normalization of the spectrum is set by the RH
sneutrino mass, with $M_N \simeq 2\times 10^{13}$ GeV
\cite{snuoriginal}. The value of the Yukawa coupling fixes the decay
rate of the  sneutrino and therefore the final reheating $T$. In the simplest
scenario where the inflaton is the lightest RH sneutrino, we would 
need $|h_N| < O(10^{-7} - 10^{-6})$ if we want to keep $T_{RH} \leq
10^7 - 10^8$ GeV in order not to have problems with thermal production of
gravitinos \cite{gravitinos}. We remark that 
$T_{end}$ in fig. (\ref{plot1}) is the temperature associated to the
radiation energy density at the end of
inflation due to dissipative effects, but this is not necessarily the
reheating $T$ typically defined as the $T$ at which the inflaton
completely decays and the Universe becomes radiation dominated. In the
cold inflation scenario the radiation energy density at the end of
inflation is always subdominant, and then reheating would proceed as
usual by the subsequent decay of the sneutrino. 

On the other hand, for a coupling $10^{-3} < |h_N| < 10^{-1}$
inflation takes place in the weak dissipative regime, which would
require a sneutrino mass of the order of $10^{12}-10^{11}$ GeV in order to
fit the COBE amplitude of the spectrum. We notice that for these
coupling and mass values, $|h_N| > 10^{-3}$ and $M_N < 10^{13}$
GeV, the adiabatic-Markovian approximation 
Eq. (\ref{markovian}) holds.  Still, the field values are  
larger than $m_P$. Again, the energy density in radiation is not
dominant at the end of inflation, and $T_{end}$ is not necessarily the
final $T_{RH}$. However, given that now we have a larger value of the
coupling $|h_N|$, the standard estimation of $T_{RH}=(90/(\pi^2
g_*))^{1/4} \sqrt{\Gamma_N m_P}$ in this regime
would give a value $O(10^{12}\, GeV)$,  beyond the 
gravitino constraint\footnote{This bound does not apply if gravitinos
  are the lightest stable SUSY particles \cite{bolz}, like in gauge
  mediated susy breaking  models \cite{gmm,fuji}.  Also the constraint
  can be relaxed for very massive gravitino, like for example in
  anomaly mediated susy breaking models \cite{anomaly}.}.

More interesting is the strong dissipative regime with  $r >
O(100)$ for $|h_N| \agt 0.1$. In this regime field values are always kept
below the cut-off scale $m_P$, which render the theory more attractive
from the point of view of particle physics. The model can be
considered as an effective model valid below the cut-off scale $m_P$,
without the need of worrying about sugra corrections. Those are kept
negligible for field values below Planck. The sneutrino mass value
varies between $10^{10}$ GeV and $2\times 10^{5}$ GeV, decreasing with the
value of the coupling. In particular, using Eqs. (\ref{pphir}) and
(\ref{spectrumr}), the amplitude of the spectrum of primordial
curvature perturbation is given by,
\be
P_{\cal R}^{1/2}\simeq \frac{1}{2} \left(\frac{15}{64
  g_*}\right)^{1/8} (4 N_e)^{3/4} C_\Upsilon^{3/8}
\left(\frac{M_N}{m_P}\right)^{3/8} \,,
\label{eqspectrum}
\ee
and using  $P_{\cal R}^{1/2}=5 \times 10^{-5}$, (and $g_*=228.75$, the
number of effective degrees of freedom for the MSSM) we have for $|h_N|
\agt 0.1$  
\be
|h_N| \simeq 8.2\times 10^{-2} \left(\frac{10^{10} GeV}{M_N} \right)^{1/3} .
\label{eqhN}
\ee
This equation summarizes the constraint on the
coupling and the sneutrino mass in order to have the strong
dissipative regime\footnote{The value of $|h_N|$ in Eq. (\ref{eqhN})
depends on the value of the top Yukawa coupling as $|h_N|
  \propto Y_t^{-1/3}$. For example, for $Y_t \simeq 1/\sqrt{4 \pi}$ we
get $|h_N| \simeq 0.2$ for $M_N = 10^{10}$ GeV.}. The larger the 
coupling $|h_N|$ is, the lighter the RH sneutrino. 

In this regime inflation ends when the energy density in
radiation becomes comparable to that of the sneutrino
field and the Universe becomes radiation dominated. Therefore, in this
case $T_{end}\simeq T_{RH}$, with values that are still larger than the
gravitino bound. Another question is about leptogenesis in this
scenario. One of the nice and more appealing features of sneutrino
inflation  is the possibility of relating in principle different
pieces of physics  like inflation, and neutrino masses and leptogenesis,
through the physics of the RH neutrinos and their couplings. 
The lepton asymmetry $Y_L=n_L/s$ is generated by the
out-of-equilibrium decay of the RH sneutrinos, and then reprocessed to
the $B-L$ asymmetry 
by sphaleron processes at a temperature around $T\sim 100$ GeV,
generating the observed baryon asymmetry $Y_B=n_B/s \simeq (8.7\pm
0.4)\times 10^{-11}$ \cite{WMAP}.  
Successful thermal leptogenesis, with the initial RH sneutrino
abundance produced out of the thermal bath, requires \cite{epsilonbound} 
$T_{RH} \geq 2 \times 10^{9}$ GeV, which is fulfilled for $|h_N| \alt
1.8$. It also requires a similar bound for the sneutrino mass, $M_N >
2\times 10^{9}$ GeV,  although the sneutrino
dominating during inflation need not necessarily be the
one originating the lepton
asymmetry. We could have for example the lighter one with the larger
Yukawa coupling as the inflaton, and the next-to-lightest being
responsible for $Y_L$. Nevertheless, there are models where these bound
can be evaded \cite{raidal}.

Before closing this section, we comment on the
predictions for the spectral index $n_S$, and the running of the spectral
index $d n_S/d \ln k$, of the primordial spectrum. Those do not vary
significantly from one regime to another. In the case of standard cold
sneutrino inflation, we have $n_S-1 \simeq -4/(2 N_e +1) \simeq -0.03$
and $d n_S/d \ln k \simeq -32/(4 N_e +2)^2 \simeq -5 \times 10^{-4}$, 
whereas in the strong dissipative regime we have  $n_S-1 \simeq -3/(2
N_e) \simeq -0.025$ and $d n_S/d \ln k \simeq -26/(4 N_e)^2 \simeq
-4.5 \times 10^{-4}$. The distinctive prediction comes from the
tensor-to-scalar ratio $r_T= P_T/P_{\cal R}$, with the primordial
spectrum of the tensor modes being $P_T\simeq 2(H/2 
\pi m_P)^2$. Whereas in cold sneutrino inflation, given that the field is
larger than $m_P$, we have \cite{csnu,snuoriginal} $r_T \simeq 16
\epsilon\simeq 0.16$, in the strong dissipative regime that ratio is
highly suppressed, with
\be
r_T \simeq 0.22 \left(\frac{0.01}{|h_N|} \right)^{12} \,.
\ee   
Future CMB experiment like Planck \cite{planck}, and also
gravitational wave detectors  currently under study
\cite{graviwaveback}, are expected to 
reach a sensitivity for $r_T$ below 0.01.  
Therefore, the lack of a signal for the primordial spectrum of
gravitational waves in future experiments will rule out sneutrino
inflation in its more standard version, but not warm sneutrino
inflation. 

\section{Lowering the post-inflation temperature}
\label{sect5}

Taking into account a {\it lighter} sneutrino from the very
beginning, there is a simpler alternative that allows to lower the
reheating $T$ at the end of warm inflation with strong dissipation, 
and which at the same time is compatible with generating the right
level of baryon asymmetry by non-thermal leptogenesis.  
Let us denote by $\phi_2$, $M_2$ the field and mass parameter of the
RH sneutrino dominating the energy density during inflation, and
$\phi_1$, $M_1$ those of a lighter RH sneutrino. 
The dominant contribution during inflation is given by $\phi_2$, 
$V\simeq M_2^2 \phi_2^2/2$, but still the lighter sneutrino can
follow a slow-roll trajectory during inflation for field values below
$m_P$, with the slow-roll parameter for $\phi_1$ being 
\bea
\eta_1 &\simeq& \frac{M_1^2}{M_2^2} \frac{2 m_P^2}{\phi_2^2}\,,\\
\epsilon_1 &\simeq& \eta_1\left( \frac{\phi_1}{\phi_2}\right)^2\,.
\eea
In order to have $\eta_1 <1$, $\epsilon_1<1$, we only need to assume
that the field  values during inflation are comparable, $\phi_1 \simeq
\phi_2< m_P$, and require $M_1/M_2 < (\phi_2/m_P)_{end} 
\simeq 1/\sqrt{2r}$, which in terms of the coupling reads:
\be
\frac{M_1}{M_2} \alt 1.6 \times 10^{-5} |h_{N2}|^{-3} \,.
\label{ratio12}
\ee

Having a second slow-rolling field does not change the primordial
spectrum during or after inflation, so the estimation given in
Eq. (\ref{eqspectrum})  applies, and the spectrum is dominated by
thermal effects. Moreover, it does not matter what is the amplitude of
the curvature perturbation generated by $\phi_1$ during inflation, by
the end of inflation it has leveled to that of $\phi_2$. The
constraint on the sneutrino mass dominating during inflation, $M_2
\alt 10^{10}$ GeV, obtained in the previous section still applies. 
      
During inflation the lightest sneutrino $\phi_1$ energy density is
subdominant. When inflation ends for $\phi_2$ it does so for
$\phi_1$. Then, this field, weakly coupled, starts oscillating and its
energy density on average behaves like 
matter. Therefore, if its decay rate is small enough, it will end up
dominating over the radiation energy density dissipated by $\phi_2$
during warm inflation. Later the field decays, and it is at this point that we
define the final reheating $T$. 
Thus, the inflationary period is controlled
by $\phi_2$, but the reheating phase is controlled by $\phi_1$, with
\be
T_{RH} \simeq \left(\frac{90}{\pi^2 g_*}\right)^{1/4} \sqrt{\Gamma_1
  m_P} \,,
\ee
where $\Gamma_1 \simeq \frac{|h_{N1}|^2}{16 \pi} M_1$, and then
\be
|h_{N1}| \simeq 0.86 \times 10^{-6}\times
\left(\frac{T_{RH}}{10^6\,GeV}\right)\left(\frac{10^8\,GeV}{M_1}\right) \,.
\ee
Leptogenesis can now proceed through the out-of-equilibrium decay of
the lightest sneutrino $\phi_1$ during the reheating period. Any
previous lepton asymmetry would be diluted by the entropy produced by
the $\phi_1$ decay. The lepton asymmetry at the end of reheating is
then given by \cite{snuoriginal,snusugra},
\be
\left.\frac{n_L}{s}\right|_{RH} \simeq |\epsilon_1| \frac{3 T_{RH}}{4 M_1} \,,
\label{nL}
\ee
with $\epsilon_1$ being the CP asymmetry generated by the decay, given
by the interference of the tree level with the one-loop amplitude,  
\be   
|\epsilon_1| \simeq \frac{3}{8 \pi (h_N^\dagger h_N)_{11}} \sum_{i\neq
  1} Im[(h_N^\dagger h_N)_{1i}]^2 \frac{M_1}{M_i} \,. 
\ee
The asymmetry parameter is bounded by\footnote{The bound is given by
  $|\epsilon_1| \leq 3 (m_3-m_1) M_1/(8 \pi v^2)$, with $m_{3,1}$
  being the light neutrino masses. We have taken for simplicity $m_1 <
m_3 \simeq \sqrt{\Delta m^2_A}\simeq O(0.05)$ eV, although we could
have for example $m_1 < m_3 \approx O(1)$ eV.} \cite{epsilonbound}
\be 
|\epsilon_1| \leq \frac{3}{8 \pi} \sqrt{\Delta m^2_A}
\frac{M_1}{v_u^2} \simeq 2 \times 10^{-8} \left(\frac{M_1}{10^8\, GeV}
\right)\,,
\label{epsilon1}
\ee
where $\Delta m^2_A\simeq (1.3 - 4.2)\times 10^{-3}$ eV$^2$ is the
the atmospheric neutrino mass squared difference
\cite{atmospheric,pdg}. From Eqs. (\ref{nL}) and (\ref{epsilon1}) we have
then the bound \cite{csnu,shafinL}, 
\be
\left.\frac{n_L}{s}\right|_{RH} \leq |\epsilon_1| \frac{3 T_{RH}}{4
  M_1} \simeq 1.5 \times 10^{-10} \frac{T_{RH}}{10^6 \, GeV} \,,
\ee
and the baryon asymmetry:
\be
\left.\frac{n_B}{s}\right|_{RH} \leq
\frac{8}{23}\left.\frac{n_L}{s}\right|_{RH} \simeq 
  5 \times 10^{-11} \frac{T_{RH}}{10^6 \, GeV} \,,
\label{YB}
\ee
and hence $T_{RH} \geq 2 \times 10^{6}$ GeV in order to match the
observed baryon asymmetry. 
On the other hand, in order to ensure the out-of-equilibrium decay of
$\phi_1$ and avoid thermal washout of the asymmetry, we require
$M_{N1} \geq T_{RH}$. Using Eqs. (\ref{eqhN}) and (\ref{ratio12}),
the limiting value \cite{csnu} $T_{RH} \simeq M_{N1} \simeq 10^6$
GeV is reached for $|h_{N2}| \alt 0.24$. 

Therefore, we have a narrow window of values $0.1 \alt |h_{N2}| \alt
0.24$, for which inflation happens in the strong dissipative regime
with $M_{N2} \simeq 10^{10}-10^{9}$ GeV, but reheating with $T_{RH}
\simeq 10^6$ GeV and {\it non-thermal} leptogenesis is given by the decay of
a lighter sneutrino with parameters $5\times 10^8 {\rm GeV} \agt
M_{N1} \agt 10^6$ GeV and $2 \times 10^{-7} \alt |h_{N1}| \alt 8.6
\times 10^{-5}$. For having warm inflation with a larger Yukawa
coupling $h_{N2} \agt 0.24$, the second lighter and long-lived
sneutrino with $M_{N1}< 10^6$ GeV can lower the final reheating $T$
below the gravitino bound, but it does not seem consistent with
non-thermal leptogenesis as we cannot satisfy at the same time $M_{N1}
\agt T_{RH}$ and $T_{RH} \simeq 10^6$ GeV. It would remain to check
whether thermal leptogenesis could be viable during the reheating period
in this case, for which one would need to set and study the Bolztman
equations describing the evolution of the different number densities,
which is beyond the scope of this paper.     

\section{Warm inflation and light neutrino masses}
\label{sect6}

In this section we briefly want  to comment on the issue of light
neutrino masses with a not too heavy sneutrino $M_{N} \leq 10^{10}$
GeV but large Yukawa couplings $|h_N| \geq 0.1$.  
Over the recent years, different neutrino experiments have
established the existence of neutrino 
oscillations driven by nonzero neutrino masses and neutrino mixing
\cite{petkov}. Atmospheric neutrino oscillation parameters read: 
\be
|\Delta m_A^2| = (1.3 - 4.2) \times 10^{-3} eV^2 \,,\;\;\; \sqrt{|\Delta
  m_A^2|} \simeq 0.05 eV \,,\;\;\; \sin  2\theta_A \geq 0.85 \,, 
\ee
while solar neutrino oscillation parameters lie in the low-LMA (large
mixing angle) solution with:
\be
|\Delta m_{sun}^2| = (8.5 - 7.4) \times 10^{-5} eV^2 \,,\;\;\; \tan
  \theta_{sun} = 0.4 ^{+0.09}_{-0.07}\,.
\ee
On the other hand, a combined analysis of the solar neutrino, CHOOZ
and KamLAND data  gives $\sin^2 \theta_{13} < 0.055$. An upper limit
on the absolute value of the masses is obtained from WMAP data as $\sum_j
m_j < (0.7 - 2.0)$ eV.

Given the superpotential Eq. (\ref{snumssm}), light neutrino masses
are given by diagonalizing the
 see-saw mass matrix \cite{seesaw}:  
\be
m_{LL} = v_u^2 h_N M_{RR}^{-1} h_N^T = U_{\nu L} diag(m_1, m_2,m_3)
U_{\nu L}^T\,,
\label{mLL}
\ee
where\footnote{Strictly speaking we have $\langle H_u \rangle = v \sin
  \beta$, wit $v=174$ GeV and $\tan \beta= \langle H_u \rangle/
  \langle H_d \rangle$. We are setting $\sin \beta \approx 1$ for
  order of magnitude estimations.} $v_u = \langle H_u \rangle \sim
174$ GeV, $m_i$ the light  LH neutrino masses, and $U_{\nu L}$ the
rotation matrix. In the Yukawa 
matrix $h_N$, each column define a vector ${\bf h_{Ni}}$ with modulus
$|h_{Ni}|$ as given in Eq. (\ref{yukawahN}). In the eigenmass basis
for the RH we have:
\be
Tr \,m_{LL} = \sum_{i=1,2,3} m_i \simeq v_u^2 \sum_{i=1,2,3}
\frac{{\bf h_{Ni}\cdot h_{Ni}}}{M_{Ni}} < O(1) {\rm eV}\,.
\ee 
For the parameters of the strong dissipative regime we clearly
exceed the WMAP bound, with ${\bf h_{N2}\cdot h_{N2}}/M_{N2} > O(30)$ eV. 

However, this applies when  assuming  a diagonal mass matrix
$M_{RR}$ in Eq. (\ref{mLL}). We can work instead with \cite{inverted}
(see also \cite{raidal})
\be
M_{RR} = \left( \begin{array}{ccc} 0 & M_{N2} &0 \\ M_{N2} & 0 & 0 \\ 0 &0
  &M_{N1} \end{array} \right) \,,   
\ee
where  
\be
Tr\, m_{LL} = \sum_{i=1,2,3} m_i \simeq v_u^2 \left(
\frac{ 2{\bf h_{N3}\cdot h_{N2}}}{M_{N2}} + \frac{{\bf h_{N1}\cdot
    h_{N1}}}{M_{N1}}\right)  < O(1) {\rm eV}\,, 
\ee 
such that the large contribution coming from the large Yukawa coupling
$h_{N2}$ can be canceled out by choosing an appropriate smaller
coupling $h_{N3} \ll h_{N2}$. This kind of scheme gives rise to light
neutrino masses with an inverted hierarchy, $m_1^2 \approx m_2^2 \gg
m_3^3$. For example, taking  ${\bf h_{N1}\cdot h_{N1}}/M_{N1} \ll {\bf
  h_{N3}\cdot h_{N2}}/M_{N2}$. 
We have then 2 almost degenerate light neutrino masses, with $|m_1|
\simeq |m_2| \simeq \sqrt{|\Delta m_A^2|}$, and a
massless one $m_3\simeq 0$, (small corrections from the Yukawas
$h_{N1}$ gives a non-zero $m_3$ value). Atmospheric neutrino
oscillations are given  by oscillations among ``13'' and ``23'', while 
solar data is 
explained by the oscillation between ``12'', with  $m_1^2 - m_2^2
\simeq \Delta m_{sun}^2$ \cite{inverted}.         

The mass parameter $M_{N1}$ can be larger or smaller than $M_{N2}$
as far as light neutrino
masses are concerned. Nevertheless, if we choose the hierarchy $M_{N2}
> M_{N1}$, the asymmetry parameter corresponding to the decay of the lightest
sneutrino is given by: 
\bea
|\epsilon_Y| &\simeq & \frac{3}{16 \pi (|h_{N1}|^2)} Im[ ({\bf h_{N3}^*
    \cdot h_{N1}})({\bf h_{N2}^*\cdot h_{N1}}) ]\frac{M_{N1}}{M_{N2}}
    \nonumber \\  
&\simeq& \frac{3}{8 \pi } \sqrt{|\Delta m_A^2|} \frac{M_{N1}}{v_2^2} \,,
\eea
where in the second line we have just assumed that there is no
hierarchy among the different components of ${\bf h_{N1}}$ 
and that phases are such that
$Im[]$ is maximal and we saturate the upper bound on the asymmetry
parameter.

\section{Conclusion}
\label{sect7}

The most important new feature for sneutrino
inflation found in this paper is a model in which
inflation is driven by just a single sneutrino field
that creates a monomial inflationary potential,
similar to chaotic sneutrino inflation
\cite{snuoriginal,snusugra,csnu}, but with the key 
difference that in the model of this paper the
inflaton amplitude is below the Planck scale.
For particle physics model building, this is an
important feature, since this model is then not susceptible
to large effects from higher dimensional operators. In particular, it
can be embedded in a sugra potential even with minimal Kahler
potential for the fields, with the exponential sugra correction in
front of the potential remaining small and under control. 
The chaotic inflation scenario of the cold inflation picture
has always been attractive for its simplicity, since it requires
just a monomial potential to realize inflation.
However the downside of this model for model building has
been that the inflaton field amplitude necessary
for inflation must be larger than the Planck scale, thus making
the model highly susceptible to higher dimensional
operator corrections. Now, taking into account dissipation, this simple
model with a monomial potential can be regarded as an effective model truly 
valid below the cut-off scale $m_P$.  

Before the results of this paper, the simplest sneutrino
model that could maintain field amplitudes below the Planck
scale was a version of hybrid inflation \cite{hsnu}.
But this model is more contrived since it requires
additional fields aside from the RH neutrino fields.
Thus the model of this paper is the simplest
realistic realization of sneutrino inflation, in the sense
that it is the minimal SUSY extension of the Standard
Model to incorporate supersymmetry and RH neutrinos,
it requires no additional fields beyond these to realize inflation,
and all higher dimensional operators which inevitably also exist
are all suppressed.

The key feature of our model that allowed the inflaton amplitude below
the Planck scale with a monomial inflation potential
was the presence of large dissipation in the inflaton
evolution equation.  Moreover as shown in Sect. \ref{sect3},
the origin of these dissipative effects arise automatically
at a first principles level for this model of
RH neutrinos coupled to the MSSM.
Thus we believe the model in this paper has several attractive
features for building a complete model that is able to describe 
both particle
physics and cosmology.

\begin{acknowledgments}
We thank Steve King for helpful discussions.
AB was funded by the United Kingdom Particle Physics and
Astronomy Research Council (PPARC).
\end{acknowledgments}


\begin{thebibliography}{99}

\bibitem{csnu} J. R. Ellis, M. Raidal, and T. Yanagida,
Phys. Lett. B {581}, 9 (2004)

\bibitem{hsnu} S. Antusch, M. Bastero-Gil, S. F. King, and Q. Shafi,
Phys. Rev. D{\bf 71}, 083519 (2005).
 
\bibitem{snuoriginal} H. Murayama, H. Suzuki, T. Yanagida, and
  J. Yokoyama, Phys. Rev. Lett. {\bf 70}, 1912 (1993). 

\bibitem{snusugra} H. Murayama, H. Suzuki, T. Yanagida, and
  J. Yokoyama, Phys. Rev. {\bf D50}, 2356 (1994). 
                                                      
                         
\bibitem{seesaw} For a review see for example: S. King,
  Rep. Prog. Phys. {\bf 67}, 107 (2004), and references therein.

\bibitem{br05} A. Berera and R. O. Ramos, Phys. Rev. D {\bf 71},
023513 (2005); Phys. Lett. B {\bf 607}, 1 (2005).

\bibitem{br} A. Berera and R. O. Ramos,
Phys. Rev. D{\bf 63} (2001) 103509;
Phys. Lett. B{\bf 567} (2003) 294.


\bibitem{wi} A. Berera,  Phys. Rev. Lett. {\bf 75} (1995) 3218;
Phys. Rev. D{\bf 54} (1996) 2519;
Phys.\ Rev.\  D{\bf 55} (1997) 3346.

\bibitem{bb1} M. Bastero-Gil and A. Berera, Phys. Rev. D {\bf 71},
063515 (2005).
            
\bibitem{mtop} P. Azzi et al., hep-ex/0404010; J. F. Arguin et al.,
  hep-ex/0507006. 
     
\bibitem{Hall:2004zr}
L.~M.~H.~Hall and I.~G.~Moss,
Phys.\ Rev.\ D {\bf 71}, 023514 (2005).

                                                               
\bibitem{oldi} A. H. Guth, Phys. Rev {\bf D23} (1981) 347;
K. Sato, Phys. Lett. B{\bf 99} (1981) 66.
                                                                                
\bibitem{ni} A. Albrecht and P. J. Steinhardt, Phys. Rev. Lett.
{\bf 48} (1982) 1220; A. Linde, Phys. Lett. {\bf 108B} (1982) 389.
                                                                                
\bibitem{ci} A. Linde, Phys. Lett. {\bf 129B} (1983) 177.

\bibitem{Guth:ec}
A.~H.~Guth and S.~Y.~Pi,
Phys.\ Rev.\ Lett.\  {\bf 49} (1982) 1110.

\bibitem{Moss:wn}
I.~G.~Moss,
Phys.\ Lett.\ B {\bf 154}, (1985) 120.

\bibitem{Berera:1995wh}
A.~Berera and L.~Z.~Fang,
Phys.\ Rev.\ Lett.\  {\bf 74} (1995) 1912.

\bibitem{Berera:1999ws}
A.~Berera,
Nucl.\ Phys.\ B {\bf 585}, (2000) 666.


\bibitem{arjunspectrum} A. N. Taylor and A. Berera, Phys. Rev. {\bf
D62} (2000) 083517.

\bibitem{hmb1}  L. M. H. Hall, I. G. Moss and A. Berera,
Phys. Rev. D{\bf 69} (2004) 083525.

\bibitem{warmspectrum} W. Lee and L.-Z. Fang, Phys. Rev. {\bf
D59} (1999) 083503; H. P. de Oliveira and S. E. Joras, Phys. Rev. {\bf
D64} (2001) 063613; J. chan Hwang and H. Noh, Class. Quantum
Grav. {\bf 19} (2002) 527  


\bibitem{warmrunning}  L. M. H. Hall, I. G. Moss and A. Berera,
Phys. Lett. B{\bf 589} (2004) 1.

\bibitem{COBE} G. F. Smoot et al., Astrophys. J. Lett. {\bf 396}
  (1996) L1; C. L. Bennet et al.,   Astrophys. J. Lett. {\bf 464}
  (1996) 1. 

\bibitem{WMAP} WMAP collab.: D. N. Spergel et al., astro-ph/0302209;
G. Hinshaw et al., astro-ph/0302217;   H. V. Peiris et al., astro-ph/0302225.

\bibitem{gravitinos}
M.~Y. Khlopov and A.~D. Linde,  Phys.
  Lett. \textbf{B138} (1984), 265--268;
J.~R. Ellis, J.~E. Kim, and D.~V. Nanopoulos,
Phys. Lett. \textbf{B145} (1984), 181; 
J.~R. Ellis, D.~V. Nanopoulos, and S.~Sarkar,
Nucl. Phys. \textbf{B259} (1985), 175; 
T.~Moroi, H.~Murayama, and M.~Yamaguchi,  Phys. Lett. \textbf{B303}
(1993), 289--294; 
M.~Kawasaki, K.~Kohri, and T.~Moroi, astro-ph/0402490. 


\bibitem{bolz} M. Bolz, A. Branderburg, W. Buch\"uller,
  Nucl. Phys. {\bf B606} (2001) 518.  
  gravitinos 

\bibitem{gmm} For a review on gauge mediation models, see for
  example:  G. F. Giudice and R. Rattazzi, Phys. Rept. {\bf 322}
  (1999) 419. 


\bibitem{fuji} M. Fuji and T. Yanagida, Phys. Lett. {\bf B 549} (2002)
  273; M. Fuji, M. Ibe and T. Yanagida, Phys. Rev. {\bf D69} (2004) 015006.

\bibitem{anomaly} T. Gherghetta, G. F. Giudice and J. D. Wells,
  Nucl. Phys. {\bf B559} (1999) 27.


\bibitem{epsilonbound}
K.~Hamaguchi, H.~Murayama, and T.~Yanagida,  Phys. Rev. \textbf{D65} (2002), 043512,
 hep-ph/0109030;
S.~Davidson and A.~Ibarra, Phys. Lett. \textbf{B535} (2002), 25--32, 
hep-ph/0202239;
G. F. Giudice, A. Notari, M. Raidal, A. Riotto, A. Strumia,
Nucl. Phys. {\bf 685} (2004) 89. 


\bibitem{raidal} M. Raidal, A. Strumia and K. Turzy\'nski,
  hep-ph/0408015. 

\bibitem{planck} http://www.rssd.esa.int/index.php?project=PLANCK

\bibitem{graviwaveback} T. L. Smith, M. Kamionkowski and A. Cooray,
  astro-ph/0506422. 

\bibitem{atmospheric} Y. Fukuda et al., Phys. Rev. Lett. {\bf 81}
  (1998) 1562; Y. Ashie et al., Phys. Rev. Lett. {\bf 93} (2004)
  101801.

\bibitem{pdg} ``Review of Particle Physics'', S. Edidelman et al.,
  Phys. Lett. {\bf B592} (2004) 1. 

\bibitem{shafinL} T. Asaka, K. Hamaguchi, M. Kawasaki, T. Yanagida,
  Phys. Lett. {\bf B464} (1999) 12; V. N. Senoguz and Q. Shafi,
  Phys. Lett. {\bf B582} (2004) 6; V. N. Senoguz and Q. Shafi,
  Phys. Rev. {\bf D71} (2005) 043514. 

\bibitem{petkov} See for example: S. T. Petkov,  hep-ph/0504166

\bibitem{inverted} R. Barbieri, L. Hall, D. Smith, A. Strumia and
  N. Weiner, JHEP 9812 (1998) 17; A. S. Joshipura, S. D. Rindani,
  Eur. Phys. J. {\bf C14} (2000) 85; R. N. Mohapatra,
  A. Perez-Lorenzana and C. A. de S. Pires, Phys. Lett. {\bf B474}
  (2000) 355; S. F. King and N. Nimai Singh, hep-ph/0007243. 

\end{thebibliography}
\end{document}